\documentclass[fleqn,10pt]{wlscirep}
\usepackage[utf8]{inputenc}
\usepackage[T1]{fontenc}
\usepackage{lineno}
\title{Individual differences in knowledge network navigation}

\author[1, 2, *]{Manran Zhu}
\author[3, 4]{Taha Yasseri}
\author[1]{János Kertész}
\affil[1]{Central European University, Department of Network and Data Science, Vienna, 1100, Austria}
\affil[2]{Corvinus University of Budapest, Center for Collective Learning, CIAS, Budapest, 1093, Hungary}
\affil[3]{University College Dublin, School of Sociology, Dublin 4, D04 V1W8, Ireland}
\affil[4]{University College Dublin, Geary Institute for Public Policy, Dublin 4, D04 V1W8, Ireland}

\affil[*]{Zhu\_Manran@phd.ceu.edu}


\begin{abstract}
With the rapid accumulation of online information, efficient web navigation has grown vital yet challenging. To create an easily navigable cyberspace catering to diverse demographics, understanding how people navigate differently is paramount. While previous research has unveiled individual differences in spatial navigation, such differences in knowledge space navigation remain sparse. To bridge this gap, we conducted an online experiment where participants played a navigation game on Wikipedia and completed personal information questionnaires. Our analysis shows that age negatively affects knowledge space navigation performance, while multilingualism enhances it. Under time pressure, participants' performance improves across trials and males outperform females, an effect not observed in games without time pressure. In our experiment, successful route-finding is usually not related to abilities of innovative exploration of routes. Our results underline the importance of age, multilingualism and time constraint in the knowledge space navigation.

\end{abstract}
\begin{document}

\flushbottom
\maketitle
\thispagestyle{empty}

\section*{Introduction}\label{sec:Introduction}

Online technologies have fundamentally changed information provision and acquisition in our societies. While, in principle, the digital information ecosystem is horizontal, easy to navigate, and egalitarian in providing access to information, in practice, the networks of information repositories have become so complex that successful navigation has turned into a real challenge~\cite{savolainen2015cognitive}. More importantly, access to information is not equally provided to all citizens: not only inequalities in access to the infrastructure, such as broadband Internet connection or smart devices are seen as privileges available to certain groups in societies, individual characteristics, such as familiarity with the technologies, digital literacy~\cite{dutton2014cultures}, education level~\cite{van2014digital}, and even personality traits~\cite{ho2005exploratory} all play a role in determining how much an individual can benefit from the open ocean of information available online~\cite{savolainen2015cognitive}. Finally, political decisions such as information sanctions or government censorship also challenge the idea of online information being "free for all" at a macro-level~\cite{gill2015characterizing}. Online information-seeking consists of search and navigation, which are two different but associated processes~\cite{mat2005associating}. To address the issue of inequality in information access, we focus on understanding how individuals navigate the knowledge space differently. This way, we can provide personalized support that caters to the specific needs of each user.

Previous research has shown that individuals exhibit distinct cognitive patterns and abilities when engaging in online information-seeking activities, with several characteristics identified as influential factors. For instance, it has been demonstrated that information-seeking performance is not solely contingent upon internet-related knowledge but is also impacted by key cognitive abilities, resulting in a disadvantage for older adults~\cite{sharit2008investigating, chevalier2015strategy}. Sex is another factor contributing to diverse information-seeking patterns, with males exhibiting higher confidence in their web navigation abilities compared to females. Females, on the other hand, tend to rely more on landmarks for navigation~\cite{mcdonald2000gender}. Individuals with different ethnic and cultural backgrounds also manifest distinct approaches to online information-seeking. A survey investigating information-seeking patterns among international and American graduate students revealed that international students prefer initiating their searches from the internet over VT E-resources such as electronic journals and databases, while American graduate students demonstrate the opposite preference~\cite{liao2007information}. Additionally, factors such as ideology~\cite{zmigrod2021cognitive} and personality~\cite{ho2005exploratory} have been shown to influence information-seeking patterns by shaping individuals' cognitive processes. Bi/multilingualism, though not directly linked to information-seeking, has been associated with various cognitive benefits~\cite{marian2012cognitive}, suggesting potential effects on information-seeking performance. Despite these insights, previous research has employed varying information-seeking tasks across experiments, and a comprehensive analysis incorporating all these factors within the same information-seeking setup is still lacking. Consequently, the relative significance of these factors in influencing information-seeking performance remains unclear.

Another motivation for our analysis stems from the connection between navigation in the physical space and knowledge space. Previous research has demonstrated that the same neural regions that are responsible for navigation in physical space are also involved in navigating the knowledge space: the hippocampus and entorhinal cortex, which contain cells that encode spatial information and enable spatial navigation, also play essential roles in other neural processes such as social cognition and memory~\cite{tavares2015map, olafsdottir2015hippocampal}. Various individual differences have been observed in spatial navigation: spatial abilities decline linearly with age~\cite{anguera2013video, newcombe2018individual}; males generally perform better than females at spatial navigation tasks~\cite{nazareth2019meta, newcombe2018individual}; and people growing up outside cities are generally better at spatial navigation~\cite{coutrot2022entropy}. Given the connections and differences between knowledge space and physical space, it is important to study if the individual differences in navigation in physical space are also present in knowledge space.

To gain insights into online navigation behaviors, researchers conducted a series of studies using Wikipedia as an observational setting~\cite{yasseri2012dynamics, yasseri2012practical, scaria2014last, takes2013mining, kattenbeck2018spatial} and utilized its well-documented network of articles as the framework for navigation studies~\cite{lamprecht2017structure, arora2022wikipedia}. The wide range of topics represented in Wikipedia (https://en.wikipedia.org/) and the platform's popularity make it a prime candidate for investigating empirical navigation behavior. In a popular online navigation game on Wikipedia, implemented in several versions such as the Wikispeedia (https://dlab.epfl.ch/wikispeedia/play/) and the Wikigame (https://www.thewikigame.com/), players try to go from one Wikipedia article (source) to another (target) through the hyperlinks of other articles within the Wikipedia website. Several navigation patterns on the Wikipedia knowledge network have been discovered: players typically first navigate to more general and popular articles and then narrow down to articles that are semantically closer to the target~\cite{west2012human}; players' search is not Markovian, meaning that a navigation step depends on the previous steps taken by the players~\cite{singer2014detecting}. When it comes to individual differences in navigation on Wikipedia, however, there is still a lack of understanding as the navigation patterns discovered so far have not taken into account personal information such as age and sex thus research has not revealed the behaviors and preferences of different demographic groups. As such, further investigations are needed to understand better how these factors may influence navigation patterns.

\begin{figure}[!htb]
\centering
\includegraphics[width=1\textwidth]{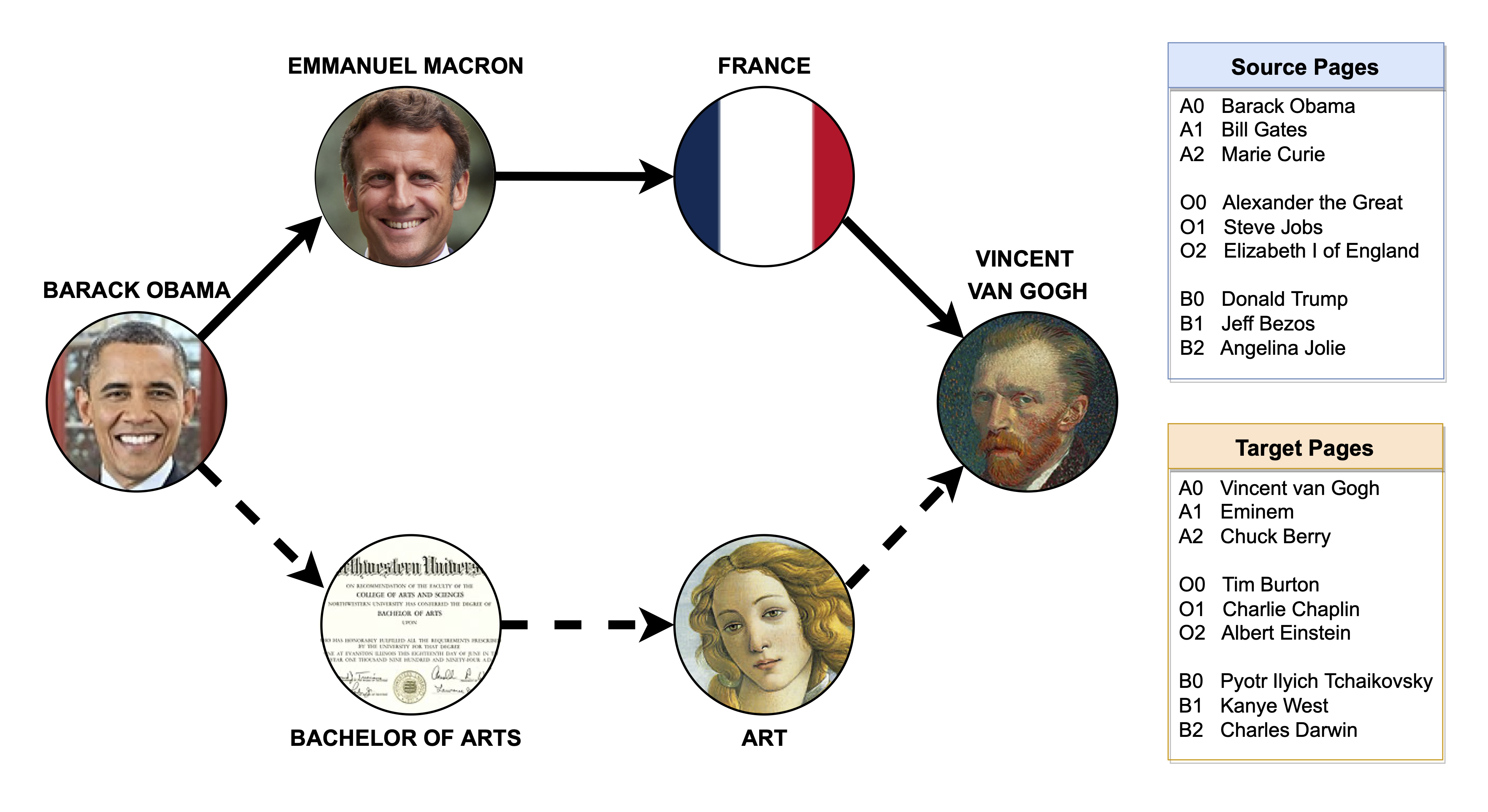}
\footnotesize
\caption{ {\bf Illustration of the Wikipedia navigation game.} In the Wikipedia navigation game, players need to go from one Wikipedia article (source page) to another (target page) through the links of other Wikipedia articles on the current page in 7 steps (Least-clicks game) or 150 seconds (Speed-race game). The figure shows two possible navigation paths from the source page \textsc{Barack Obama} to the target page \textsc{Vincent van Gogh}: 1) \textsc{Barack Obama} to \textsc{Emmanuel Macron} to \textsc{France} to \textsc{Vincent van Gogh} (solid arrows); and 2) \textsc{Barack Obama} to \textsc{Bachelor of Arts} to \textsc{Art} to \textsc{Vincent van Gogh} (dotted arrows). Participants each played nine rounds of games whose source page and target page are shown in the figure. The games are divided into three sessions A, O, and B, with three games in each session. The order of the games is randomized in each game session, and the order of sessions A and B are randomized to reduce the effect of the games' order on performance. Attributions to the images used in the figure are included in the references~\cite{fig1, fig2, fig3, fig4, fig5, fig6}.  
}
\label{fig:Illustration}
\end{figure}

To gain a better understanding of how navigation on the knowledge network is affected by individual characteristics, we conducted an online experiment where we hired 445 participants from the US to play nine rounds of Wikipedia navigation games (illustration in Fig. \ref{fig:Illustration}) and to fill in a survey afterwards about their personal information such as age, gender, and answer questions which enabled us to characterize their big five personality traits~\cite{cobb2012stability} (details in Methods). In each game, players can opt for a Speed-race or a Least-clicks challenge. To win, they must reach the target page within 150 seconds for the Speed-race games or in 7 steps for Least-clicks games. We sought to answer the question of whether individuals with certain characteristics possess an advantage over others in our navigation tasks, and if so, which are those characteristics. Moreover, using a uniqueness measure proposed in this work, we investigated if certain players are more creative than others, meaning that they not only tend to win the navigation games but also take unusual routes to the target. 

\section*{Results}\label{sec:Results}

\begin{figure}[!htb]
\centering
\includegraphics[width=1\textwidth]{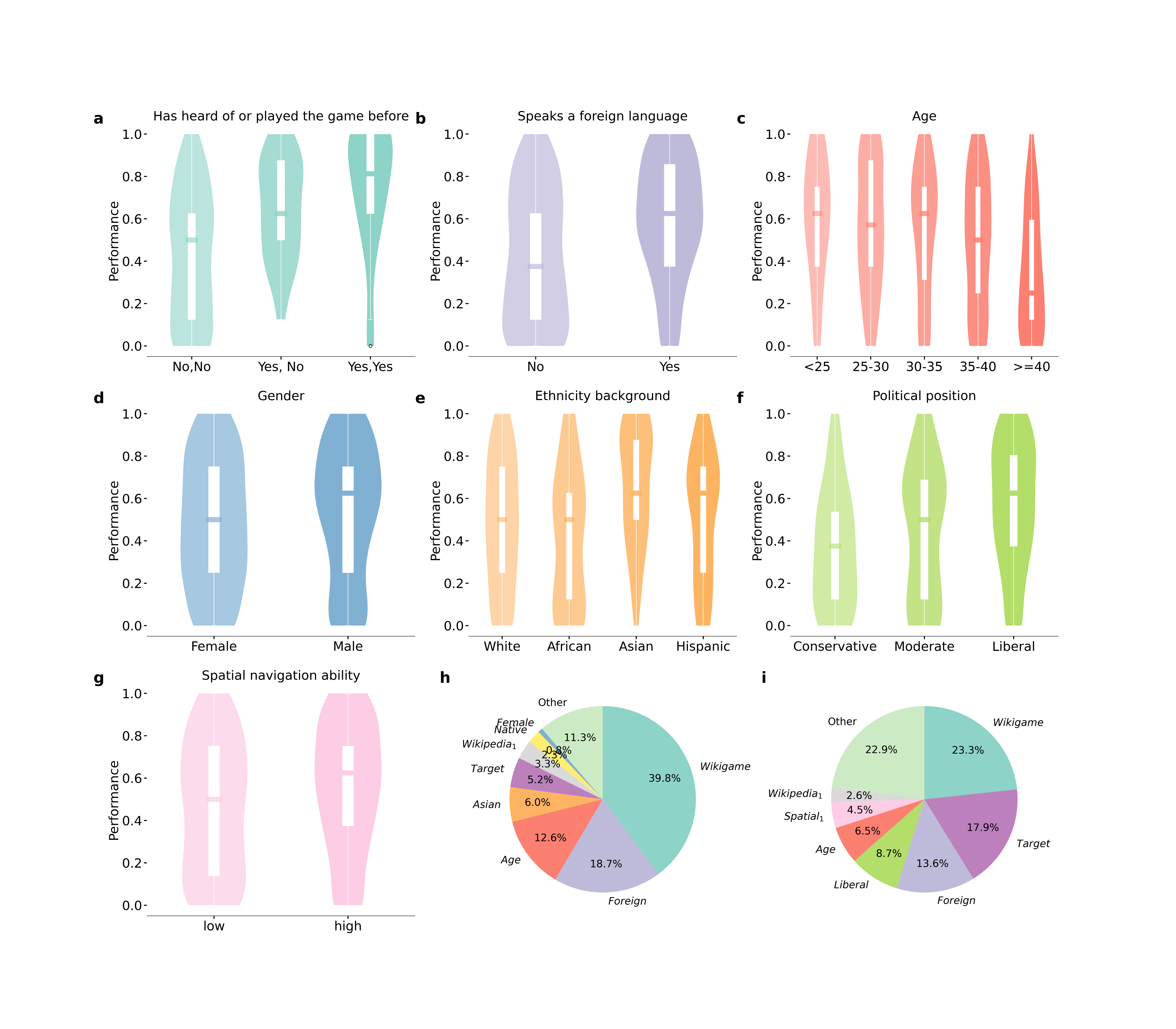}
\footnotesize
\caption{ {\bf Performance in subgroups.} The figure shows the navigation performance distribution for participants with different characteristics, where performance is measured as the ratio of games won by each participant. Fig. a-h shows the distribution of navigation performance for the participants' eight characteristics: a) age, b) gender, c) foreign language skills, d) ethnic background, e) political view, f) spatial navigation skills (the first principal component of the spatial navigation related questions $Spatial_{1}$), g) prior experience with the Wikipedia navigation game and h-i) the percentage of deviance explained by each variable as they were added as the covariate to the regression model (see details in Methods) in the Speed-race game and Least-clicks game respectively, normalized by the total variance explained by the multiple regression of all individual characteristics in Table \ref{tab:Regression}. }
\label{fig:Violinplots}
\end{figure}

\noindent\textbf{Impact of individual characteristics on navigation performance}
Upon controlling for participants' prior experience with the Wikigame, Wikipedia, and the target page, the ability to speak a foreign language (Cohen's d = 0.57) emerged as the most influential predictor of success in both game types (Table \ref{tab:Regression}). Those who are proficient in a foreign language have, on average, a 40\% higher chance of winning a game (Fig. \ref{fig:Violinplots}b and Table \ref{tab:Regression}) and it alone explains 18.7\% and 13.6\% of the total explained deviance in the Speed-race and Least-clicks games respectively (Fig. \ref{fig:Violinplots}h-i). Another key factor significantly impacting navigation performance in both game types is age (Pearson's r(395) = -0.30): younger participants exhibit significantly enhanced performance (Fig. \ref{fig:Violinplots}c and Table \ref{tab:Regression}). Specifically, age explains 12.6\% and 6.5\% of the total explained deviance in the two types of games respectively (Fig. \ref{fig:Violinplots}h-i). 

Our study reveals that prior experience with the Wikipedia navigation game, proficiency in using Wikipedia, and familiarity with the target Wikipedia page give a significant advantage to players in both game types (Table \ref{tab:Regression}). Participants who have previously played the Wikipedia navigation game are 1.7 times more likely to achieve success in our experiment than those who have not played or heard of the game before (Fig. \ref{fig:Violinplots}a). A robust understanding of the target page predicts success overall (Table \ref{tab:Regression}), and it holds greater importance as the second most significant predictor in the least clicks games compared to speed-race games (Fig. \ref{fig:Violinplots}h-i).

Regarding other individual characteristics, distinctions arise between the two types of games (see Supplementary Figure $2$ for the game time distribution of the two games). Among participants who chose to play games featuring time constraints, male participants of Asian ethnicity without native-level fluency in a foreign language tend to demonstrate stronger performance (Table \ref{tab:Regression}, Fig. \ref{fig:Violinplots}d-e). Conversely, among participants who chose to play games with distance constraints, enhanced performance is associated with having a liberal stance and reporting greater spatial navigation ability (Table \ref{tab:Regression}, Fig. \ref{fig:Violinplots}f-g). Furthermore, our findings indicate a significant performance improvement among participants in the timed games, while such improvement was not evident in the games constrained by distance (as indicated by the Order variable).

\begin{figure}[!htb]
\centering
\includegraphics[width=.8\textwidth]{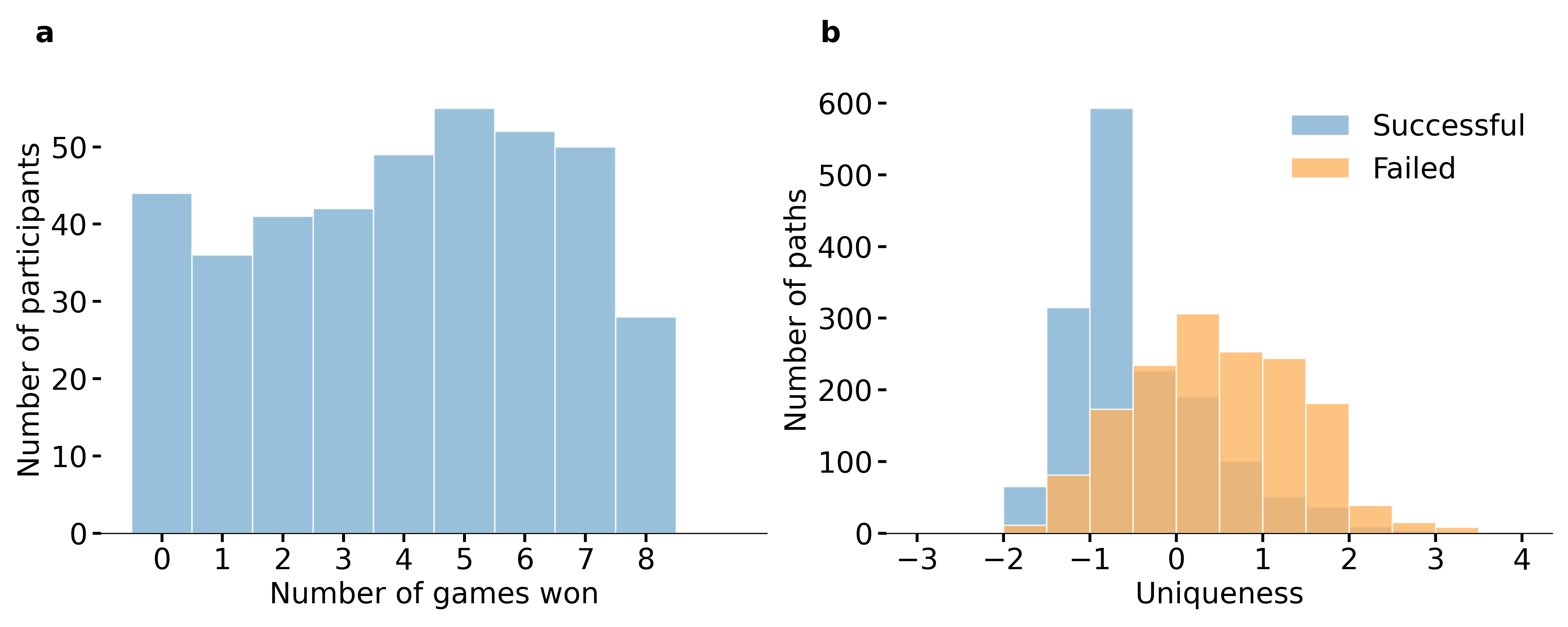}
\footnotesize
\caption{{\bf Distribution of Performance and Uniqueness.} a) shows the distribution of the total number of games the participants won in the experiment. b) shows the distribution of uniqueness scores for both successful and unsuccessful navigation paths. The uniqueness score quantifies the distinctiveness of a navigation path relative to others. For the definition and computation of the uniqueness score, refer to the Methods section. }
\label{fig:Distribution}
\end{figure}

\vspace{\baselineskip}\noindent\textbf{Interplay between success and uniqueness}
We observed considerable variation in both the success and uniqueness of participants' navigation routes: while some participants succeeded in all games, others succeeded in none (Fig. \ref{fig:Distribution}a). Additionally, some participants opted for mainstream routes, whereas others ventured onto less-traveled paths (Fig. \ref{fig:Distribution}b). Unsuccessful navigation paths exhibited, on average, greater uniqueness compared to successful paths (Fig. \ref{fig:Distribution}b). This observation arises from the fact that instances of becoming lost or deviating from the intended course not only lead to navigation failure but also to higher uniqueness scores, thus influencing the distribution of uniqueness. Figure \ref{fig:Uniqueness} visualizes the uniqueness score of the successful navigation paths for nine games respectively. 

Our regression analysis on the uniqueness scores of navigation paths reveals that akin to success, individual characteristics also influence the uniqueness of successful routes. Specifically, among participants who chose to play the games under time constraints, younger and left-handed participants (third principal component of reported spatial abilities) tend to navigate through more unique paths to reach the target (Table \ref{tab:Regression}). Conversely, among participants who chose to play the games under distance constraints, no discernible traits display a significant correlation with the degree of path uniqueness (Table \ref{tab:Regression}).

\section*{Conclusion}\label{sec:Conclusion}
Our study highlights the role of individual characteristics in participants' navigation performance within the knowledge space, with this influence being modulated by constraints such as time and distance. We discovered that prior experience with Wikipedia, the navigation game, and familiarity with the target page are significant predictors of better navigation, likely due to the nature of the game. Controlling these factors, being young and multilingual consistently predict good navigation performance irrespective of the type of constraints, indicating the fundamental role of age and multilingualism in knowledge space navigation. 

Regarding other traits, distinctions emerge when considering the two categories of constraints. Among participants who chose to play Speed-race games involving time constraints, superior performance is exhibited by male participants with an Asian ethnic background who do not speak a foreign language at a native level. Conversely, among participants who chose to play the Least-clicks games with distance constraints, enhanced performance is associated with participants identifying as liberal and self-reporting better spatial navigation skills. Notably, as the participants play more rounds of the game, they show significant improvement in Speed-race games but not in Least-clicks games. Using a uniqueness measure proposed in this work, we showed that beyond navigation success, individual traits also impact route uniqueness: under time pressure, younger and left-handed participants tend to pursue more distinct routes, unlike tasks with distance constraints.

\section*{Discussion}\label{sec:Discussion}
Previous research primarily focused on linking individual traits to navigation within physical space. Our study expands this literature by examining navigation within the knowledge space. Similar to physical space navigation~\cite{anguera2013video, spiers2021explaining}, age acts as an inhibitor here, likely due to declining cognitive abilities associated with age, impacting fluid intelligence, perceptual speed, memory, and vocabulary~\cite{ghisletta2012two}. Bilingualism has been demonstrated to have various cognitive benefits including improved executive control and protection against cognitive decline~\cite{bialystok2012bilingualism}. Our study underscores speaking a foreign language as the most potent predictor of performance in knowledge space navigation, indicating an additional cognitive advantage associated with multilingualism.

In addition, our observations indicate that individual characteristics, including sex, ethnicity, native language, political stance, and reported spatial navigation skills, significantly influence navigation performance in one type of game (with time or distance constraints) but not the other. To fully understand these effects, further investigation is necessary. One possible explanation is that these factors may be associated with other cognitive processes affecting navigation performance, which were not included in our study. Sex differences have been observed in spatial navigation tasks~\cite{lovden2007quantitative, padilla2017sex, gron2000brain}. Our finding aligns with existing research indicating that test timing moderates sex differences in spatial navigation performance~\cite{nazareth2019meta}. The impact of time pressure might be relevant to the fact that anxiety and self-doubt hinder encoding of spatial features~\cite{saucier2002sex}, with females typically reporting higher spatial anxiety and lower self-confidence than males~\cite{huang2017timing, lawton1994gender}. We have not seen any significant role of the Big Five personality traits~\cite{cobb2012stability} in our experiment, despite their expected impact on online navigation performance~\cite{ho2005exploratory}. This variation may be due to differences in the navigation tasks and the experimental environment. Based on the insights gained from our current work, we are planning a new series of experiments that will incorporate more objective cognitive measures in order to draw firmer conclusions.

Creativity, as defined conventionally, involves originality and effectiveness\cite{runco2012standard, campbell2022partial}. Our discovery extends beyond navigation success (effectiveness), revealing that route uniqueness (originality) is also influenced by individual differences, aligning with past research demonstrating frequent detour use in navigation\cite{spiers2015solving} with these detours reflecting individual traits\cite{gulyas2020role}. Traits included in our experiment that predict successful route-finding to the target usually do not necessarily correspond to innovative route-finding abilities.

Our results have far-reaching implications. When it comes to government practices of digital services, the concept of "online only" has already been challenged by scholars relying on the fact that people of certain characteristics, particularly age, are less likely to be able to get online, and therefore there must be alternatives available to them~\cite{hunsaker2018review}. While this notion is becoming more dilute as Internet penetration reaches close to 100 percent in developed countries, it is still essential to note that being online means different things for different people based on their characteristics. If something is "up on the Internet", it does not necessarily mean everyone can find it.

Our study exhibits several limitations. Firstly, to enhance the robustness of our results, we should consider integrating additional variables that measure participants' engagement, working memory, anxiety levels, and objective spatial abilities, which could potentially impact navigation performance. Secondly, it's important to note that our findings pertain specifically to a controlled navigation game scenario, where the content involves Wikipedia pages and the source and target are notable individuals. Caution should be exercised when attempting to extrapolate these findings to real-world navigation tasks within a broader knowledge domain. Thirdly, it's worth emphasizing that our research focuses primarily on knowledge navigation, which represents just one facet of the broader online information-seeking process and should be distinguished from knowledge search. Finally, given that participants can select their preferred game type, we observed significant self-selection bias for participants with different sex and reported spatial navigation skills (see Supplementary Table S$6$), which should be considered when interpreting the results. To address these limitations, we have initiated a follow-up experiment to reevaluate participants and include additional moderator variables in our survey.

Our study extends previous research on individual differences in spatial navigation to navigation in knowledge space. An intuitive next research phase could involve constructing mathematical models that integrate personal traits to elucidate participants' navigation behavior. Additionally, exploring whether and how navigation experiences can be enhanced for individuals with specific characteristics in future experiments is a viable avenue for investigation.

\section*{Methods}\label{sec:Methods}
\textbf{The experiment}
We conducted an online experiment where we hired 445 participants (397 participants after removing participants who did not finish the experiment or did not pass the attention check, and dropping data that had recording errors) from the United States on the online crowdsourcing platform Prolific (https://www.prolific.co/) to play nine rounds of the Wikipedia navigation game and fill in a survey on the survey platform Qualtrics (https://www.qualtrics.com/uk/). At the end of the experiment, each participant received a fixed rate base payment of 5 pounds and a bonus payment of 0.5 pounds for each game they won. To get a balanced population, we applied the following prescreening conditions: i) participants are from the United States, ii) an equal number of female and male participants, iii) participants with White, Asian, Hispanic, and African ethnicity consist $\sim$50\%, $\sim$17\%, $\sim$17\% and $\sim$17\% of the sample respectively. 

In the game sessions, players are given two Wikipedia pages as the source and the target in each game. To reduce the disparities in prior knowledge among the participants, the source and target pages are chosen to be similarly distanced (2 or 3 steps away on the Wikipedia network) pages about renowned individuals from various domains such as artists, directors, scientists, and politicians, spanning different historical periods and encompassing both genders. The players start from the source page and navigate to the target page by clicking on the hyperlinks to other Wikipedia articles on the page. To win each game, they should reach the target page in at most 7 steps (Least-click game) or within 150 seconds (Speed-race game). Each participant plays nine rounds of games grouped into three sessions with a one-minute break between the sessions. After the game sessions, participants first finished a 50-question Big Five personality test (https://openpsychometrics.org/tests/IPIP-BFFM/) measuring their five personality traits: openness to experience, conscientiousness, extroversion, agreeableness, and neuroticism. To control other factors that may affect navigation performance, we then asked six groups of questions about their \textit{i}) employment status, \textit{ii}) education background, \textit{iii}) spatial navigation habit, and their prior experience with \textit{iv}) the Wikipedia navigation game, \textit{v}) the Wikipedia website and \textit{vi}) computer games. Lastly, we asked participants demographic questions about their age, gender, ethnicity, political position, and language skills. See the Supplementary Material for a complete list of the questions in the survey. One of the games with the source page "Alexander the Great" and target page "Tim Burton" turned out to be much more difficult than the other games ($> 3\sigma$), and is therefore counted as an outlier and excluded from our analysis. After the exclusion, the eight rounds of navigation tasks reached a Cronbach's alpha score of 0.76, indicating fair internal reliability of the navigation task. 

\vspace{\baselineskip}\noindent\textbf{Individual characteristics}
Encoding the participants' answers to the questions in the survey (see encoding details in the Supplementary Material), we end up with 18 control variables characterizing the participants by the six groups of questions specified above, 5 control variables indicating the game, game type (Speed-race or Least-clicks), round number of the game and participants' familiarity of the source and target Wikipedia articles of the game played by each participant. In addition, we adopted 11 independent variables describing the participants' big five personality traits, age, gender, ethnic background, political position, and foreign/native language skills. To reduce the strong correlation and anti-correlation present among the control variables, we conducted a principal components analysis (PCA)~\cite{dunteman1989principal} in each question group and summarized 80\% of the variance by a reduced set of variables (principal components). The final list of the 13 control variables and their respective loadings from the original variables are shown in Table \ref{tab:Loadings}. Descriptive statistics of the participants' characteristics can be found in Supplementary Table S$1$ in the Supplementary Material. As shown, male participants in our experiment are, on average, younger and less liberal, with a more varied ethnic background. They are also more likely to speak a foreign language and have prior experience with the Wikipedia navigation game. Female participants prefer to play the navigation game without time constraints (Least-clicks game), whereas males tend to race for speed (Speed-race game). Regarding the Big Five personality score, we did not observe big differences between male and female participants (Maximum t value = 1.75).

\vspace{\baselineskip}\noindent\textbf{Navigation paths}
A navigation path of a participant refers to the sequence of Wikipedia articles, or Wikipages, clicked by the participant in a game. Representing the hyperlinking structure of the English Wikipedia as a directed graph $G = (V, E)$, with $V = \{a_{k}\}$ denoting the set of all the Wikipages $a_{k}$ and $E = \{H_{kl}\}$ denoting the set of all the existing hyperlinks $H_{kl}$ from $a_{k}$ to $a_{l}$, the navigation path with $N$ steps for the $n$th participant in the $i$th navigation game $g_{i}$ can then be represented as a sequence $P_{n}^{i} = (a_{k})_{k=0}^{N}$ on the Wikipedia graph $G$, where $i = 1, 2, ..., 8$ and $n = 1, 2, ..., 397$. Denoting the source and target Wikipages of the game $g_{i}$ by $A_{s}^{i}$ and $A_{t}^{i}$, the navigation path $P_{n}^{i} = (a_{k})_{k=0}^{N}$ for the $n$th participant in the $i$th game is successful if it starts from the source and reaches the target, i.e. $a_{0} = A_{s}^{i}$ and $a_{N} = A_{t}^{i}$, and not successful if $a_{0} = A_{s}^{i}$ and $a_{N} \neq A_{t}^{i}$. Given the navigation paths of all the participants in all the games, we measure the success of the $n$th participant in the $i$th game by a binary variable $s_{n}^{i}$, which takes the value $1$ if the navigation path $P_{n}^{i}$ is successful otherwise $0$. 

\vspace{\baselineskip}\noindent\textbf{Quantifying the uniqueness of the navigation paths}
To understand how the navigation paths differ, we first trained a 64-dimensional node embedding for each Wikipage $a_{i}$ over the English Wikipedia graph $G$ using the DeepWalk~\cite{perozzi2014deepwalk} algorithm. Graph embedding is a technique to represent each node in the graph as a numerical vector in a continuous space with similar nodes placed close to each other. This allows us to quantify the dissimilarity of two nodes as the distance between the respective vectors. Our graph embedding assigns a 64-dimensional numerical vector $\vec v_{i}$ to each Wikipage $a_{i}$, using which we constructed a semantic distance measure between the pairs of the Wikipages:

\begin{equation}
    d(a_i, a_j) = 1 - \frac{\vec v_i \cdot \vec v_j}{\|\vec v_i\|\|\vec v_j\|},
\end{equation}

\noindent where the semantic distance $d(a_i, a_j)$ between the Wikipages $a_i$ and $a_j$ is defined as the cosine distance between their graph embeddings $\vec v_i$ and $\vec v_j$. To evaluate the performance of the embedding, we tested it over the WikipediaSimilarity 353 Test~\cite{witten2008effective}, which is an adoption of an earlier dataset, WordSimilarity 353 Test~\cite{finkelstein2001placing}, for measuring semantic relatedness among words. Our graph embedding gives a Spearman rank correlation score of 0.667 with the WikipediaSimilarity 353 test, a performance comparable to the state-of-the-art semantic relatedness measure of Wikipedia pages~\cite{singer2013computing}. 

\begin{figure}
    \centering
    \includegraphics[width=.8\textwidth]{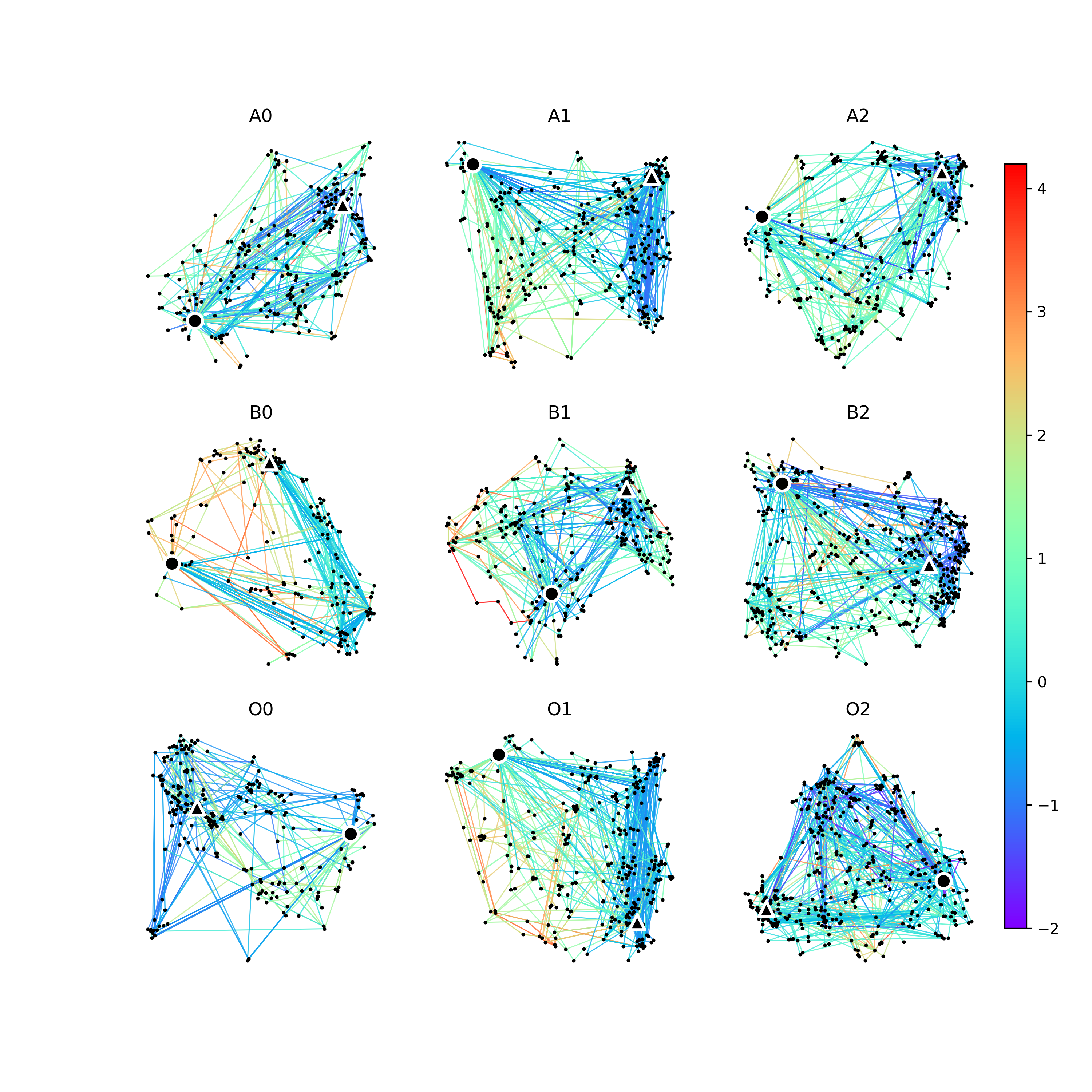}
    \caption{\textbf{Visualization of the uniqueness scores of the navigation routes} The figure displays uniqueness scores for successful navigation paths across nine distinct games. The black circle and triangle denote the source and target Wikipedia pages of the respective game, and black dots represent visited Wikipedia articles, and their positions reflect their new two-dimensional coordinates derived from reducing the original 64-dimensional embeddings using the TSNE technique~\cite{van2008visualizing}. Lines indicate successful navigation paths within the games, with line color corresponding to the uniqueness scores of these paths. }
    \label{fig:Uniqueness}
\end{figure}

Given two navigation paths $P_{m}^{i} = (a_{k})_{k=0}^{M}$ and $P_{n}^{i} = (a_{l})_{l=0}^{N}$ of the $m$th and $n$th participants in the game $g_i$, we define the distance between the two paths as the Hausdorff distance~\cite{besse2015review, rockafellar2009variational} between the two sets of Wikipages:

\begin{equation}
    D_{H}(P_{m}^{i}, P_{n}^{i}) = Max\{\sup_{a_k \in P_{m}^{i}}d(a_k, P_{n}^{i}), \sup_{a_l \in P_{n}^{i}}d(P_{m}^{i}, a_l)\}
\end{equation}

\noindent where $d(a_k, P_{n}^{i}) = \inf_{a_l \in P_{n}^{i}} d(a_k, a_l)$ quantifies the distance from the Wikipage $a_k$ to the navigation path $P_{n}^{i}$, defined as the smallest distance from $a_k$ to any Wikipage $a_l$ in $P_{n}^{i}$. Given the distance between any two navigation paths, we defined the uniqueness of a successful navigation path $P_{n}^{i}$ of the $n$th participant in the game $g_i$ as its average distance to all the other successful navigation paths in the same game:

\begin{align}
    u_{n}^{i} &= \frac{1}{K_{i} - 1} \sum_{s_{m}^{i}=1, m \neq n}^{} D_{H} (P_{n}^{i}, P_{m}^{i})\\
    \Tilde{u}_{n}^{i} &= \frac{u_{n}^{i} - \mu}{\sigma}  
\end{align}

\noindent where $K_{i}$ is the total number of successful navigation paths in the $i$th game. Standardizing the uniqueness of the navigation paths by the average uniqueness score $\mu$ and standard deviation $\sigma$ of all the successful navigation paths within the game, we get the standardized uniqueness $\Tilde{u}_{n}^{i}$ for the $n$th participant in the $i$th game. A visualization of the uniqueness scores for each successful navigation path is shown in Fig. \ref{fig:Uniqueness}. 

\vspace{\baselineskip}\noindent\textbf{Regression models}
To investigate the impact of individual characteristics on navigation success and creativity, we employed four regression models. For navigation success, we conducted separate logistic regression analyses for games with time and distance constraints. The dependent variable was the binary measure $s_{n}^{i}$ of successful or unsuccessful navigation for the $n$th participant in the $i$th game. Creativity in navigation was assessed using linear regression for each game type, with the standardized uniqueness score as the dependent variable. Ethnicity was represented as two binary variables indicating Asian/African American identity and political orientation was captured as a binary variable indicating liberal stance, because being Asian/African American and liberal are significant (p < 0.01) predictors of navigation performance while other categories of ethnicity and political positions are not. As control variables, we included a dummy variable representing the index of the eight games to account for differing difficulties, and a numeric variable indicating the order of the game to control for attention changes during the experiment. The final regression results are presented in Table \ref{tab:Regression} (the dummy variables indicating the game index were not shown in the table for visualization simplicity. For the full regression results, see Supplementary Table S$4$). The correlation among predictors significantly associated (p < 0.01) with navigation performance is illustrated in Supplementary Figure $1$, and the variance inflation factors (VIF) of all the independent variables are shown in Supplementary Table S$5$. The low VIF values (Max=2.38) indicate that the collinearity issue of our model is negligible. To test if the main effects of navigation performance in our models are still valid when interactions among the independent variables are considered, we conducted two extra logistic regressions where the interactions are included (see Supplementary Table S$4$ for details on how the interaction terms were selected). The regression results and the VIFs of the independent variables for all the logistic regression models are shown in Supplementary Tables S$4$-$5$. As demonstrated, the significant predictors of navigation performance maintain significance in both Speed-race and Least-clicks games after the inclusion of interaction terms (except Wikipedia\textsubscript{1} for the Least-clicks games, which remains significant at p < 0.05 after introducing the interaction terms). While certain variables, like employment status, computer games proficiency, and interaction terms, have achieved significance, their influence on the main effects observed is minimal. Therefore our primary focus in this study centers on the main effects of personal characteristics. To assess the different impact of each predictor on navigation performance, we also conducted a series of logistic regressions for Speed-race and Least-clicks games respectively where we added predictors one by one and presented the regression outcomes in Supplementary Table S$2-3$.

\begin{table}[!htbp] \centering 
  \caption{\textbf{Regression models for the success and uniqueness of the navigation routes} The table presents logistic regression results for navigation route success (first two columns) and linear regression results for route uniqueness (last two columns) in Speed-race and Least-clicks games. Coefficients are highlighted in bold when their corresponding variables significantly predict the dependent variable (p < 0.01). The dummy variables indicating the eight games are omitted in the results for simplicity. } 
  \label{tab:Regression} 
\begin{tabular}{@{\extracolsep{5pt}}lcccc} 
\\[-1.8ex]\hline 
\hline \\[-1.8ex] 
\\[-1.8ex]  & \multicolumn{2}{c}{\textit{Dependent variable: Success}} & \multicolumn{2}{c}{\textit{Dependent variable: Uniqueness}} \\ 
\cline{2-3} \cline{4-5} 
\\[-1.8ex] & \multicolumn{2}{c}{\textit{logistic}} & \multicolumn{2}{c}{\textit{OLS}} \\ 
\\[-1.8ex] & Speed-race games & Least-clicks games & Speed-race games & Least-clicks games\\ [1.8ex]
\hline \\[-1.8ex]
  Age & $-$\textbf{0.053}$^{***}$ (0.008) & $-$\textbf{0.021}$^{***}$ (0.006) & $-$\textbf{0.018}$^{**}$ (0.007) & $-$0.002 (0.004) \\ 
  Female & $-$\textbf{0.362}$^{**}$ (0.136) & 0.086 (0.118) & $-$0.127 (0.100) & $-$0.087 (0.070) \\
  Asian American & \textbf{0.812}$^{***}$ (0.181) & 0.168 (0.159) & 0.215 (0.113) & 0.022 (0.089) \\ 
  African American & $-$0.452$^{*}$ (0.186) & $-$0.293$^{*}$ (0.138) & $-$0.024 (0.146) & $-$0.021 (0.085) \\ 
  Foreign Language (Native) & $-$\textbf{0.601}$^{**}$ (0.190) & $-$0.130 (0.155) & $-$0.129 (0.127) & 0.045 (0.092) \\ 
  Foreign Language & \textbf{0.721}$^{***}$ (0.141) & \textbf{0.457}$^{***}$ (0.120) & 0.156 (0.096) & 0.066 (0.072) \\ 
  Liberal & 0.086 (0.137) & \textbf{0.411}$^{***}$ (0.117) & 0.125 (0.099) & $-$0.026 (0.071) \\ 
  Agreeableness & $-$0.012 (0.015) & $-$0.002 (0.013) & 0.011 (0.011) & $-$0.002 (0.007) \\ 
  Conscientiousness & 0.026 (0.015) & 0.020 (0.013) & 0.007 (0.010) & 0.001 (0.007) \\ 
  Extroversion & $-$0.031$^{*}$ (0.014) & $-$0.013 (0.013) & 0.0005 (0.010) & $-$0.004 (0.008) \\ 
  Neuroticism & $-$0.007 (0.014) & 0.021 (0.012) & 0.001 (0.009) & $-$0.001 (0.007) \\ 
  Openness & 0.018 (0.015) & $-$0.017 (0.013) & $-$0.002 (0.010) & 0.009 (0.008) \\ 
  Wikipedia\textsubscript{1} & \textbf{0.206}$^{***}$ (0.060) & \textbf{0.130}$^{**}$ (0.050) & 0.024 (0.041) & 0.028 (0.029) \\ 
  Wikipedia\textsubscript{2} & 0.081 (0.085) & $-$0.108 (0.078) & 0.079 (0.058) & 0.028 (0.047) \\ 
  Spatial\textsubscript{1} & 0.097 (0.053) & \textbf{0.160}$^{***}$ (0.048) & $-$0.034 (0.035) & 0.017 (0.028) \\ 
  Spatial\textsubscript{2} & 0.004 (0.059) & 0.067 (0.057) & 0.014 (0.038) & $-$0.026 (0.033) \\ 
  Spatial\textsubscript{3} & 0.081 (0.068) & $-$0.111$^{*}$ (0.053) & \textbf{0.154}$^{**}$ (0.048) & $-$0.022 (0.033) \\ 
  Spatial\textsubscript{4} & $-$0.030 (0.075) & 0.120 (0.065) & 0.017 (0.048) & $-$0.049 (0.040) \\ 
  Employment\textsubscript{1} & 0.033 (0.048) & $-$0.014 (0.042) & 0.056 (0.034) & 0.001 (0.025) \\ 
  Employment\textsubscript{2} & $-$0.045 (0.058) & $-$0.107$^{*}$ (0.049) & $-$0.003 (0.039) & 0.009 (0.028) \\ 
  Employment\textsubscript{3} & $-$0.116 (0.079) & $-$0.023 (0.067) & $-$0.018 (0.058) & $-$0.036 (0.041) \\ 
  Education\textsubscript{1} & 0.063 (0.060) & $-$0.083 (0.048) & $-$0.091$^{*}$ (0.041) & 0.017 (0.029) \\ 
  Education\textsubscript{2} & $-$0.047 (0.098) & $-$0.023 (0.081) & 0.012 (0.076) & 0.063 (0.050) \\ 
  Computer\textsubscript{1} & $-$0.102$^{*}$ (0.047) & 0.051 (0.045) & 0.005 (0.034) & 0.008 (0.027) \\ 
  Computer\textsubscript{2} & $-$0.020 (0.084) & $-$0.072 (0.077) & 0.009 (0.056) & $-$0.030 (0.046) \\ 
  Prior (Wikigame) & \textbf{0.562}$^{***}$ (0.107) & \textbf{0.394}$^{***}$ (0.110) & 0.041 (0.062) & 0.022 (0.057) \\ 
  Prior (Source Page) & $-$0.087 (0.068) & $-$0.021 (0.057) & $-$0.009 (0.050) & $-$0.056 (0.034) \\ 
  Prior (Target Page) & \textbf{0.192}$^{**}$ (0.068) & \textbf{0.264}$^{***}$ (0.055) & 0.011 (0.045) & 0.016 (0.033) \\ 
  Order & \textbf{0.103}$^{***}$ (0.023) & 0.020 (0.019) & 0.028 (0.015) & 0.006 (0.011) \\ 
  Constant & $-$0.192 (0.745) & $-$0.489 (0.683) & $-$0.140 (0.512) & 0.095 (0.424) \\ 
 \hline \\[-1.8ex] 
Observations & 1,479 & 1,662 & 695 & 899 \\ 
R$^{2}$ &  &  & 0.105 & 0.040 \\ 
Adjusted R$^{2}$ &  &  & 0.056 & 0.00003 \\ 
Pseudo R$^{2}$ & 0.237 & 0.117 & & \\
Deviance & 1644.4 & 2085.3 & & \\
Null Deviance & 2045.0 & 2292.9 & & \\
Log Likelihood & $-$822.205 & $-$1,042.631 &  &  \\ 
Akaike Inf. Crit. & 1,718.410 & 2,159.261 &  &  \\ 
Residual Std. Error &  &  & 1.044 (df = 658) & 0.910 (df = 862) \\ 
F Statistic &  &  & 2.140$^{***}$ (df = 36; 658) & 1.001 (df = 36; 862) \\ 
\hline 
\hline \\[-1.8ex] 
\textit{}  & \multicolumn{4}{r}{$^{*}$p$<$0.05; $^{**}$p$<$0.01; $^{***}$p$<$0.001} \\ 
\end{tabular} 
\end{table} 

\begin{table}[!htp]
\centering
\caption{\textbf{Loadings of the principle components} The table displays encoded variables (first column) and their corresponding loadings on the primary principal components in each question category, retaining at least $80\%$ of the variance within each category. Loadings quantify the extent to which original variables contribute to specific principal components. Larger values, regardless of sign, indicate a stronger association between the original variable and the principal component. The sign of the loading indicates whether the correlation between the variable and component is positive or negative.}
\label{tab:Loadings}
\begin{tabular}{p{.2\textwidth}p{.15\textwidth}p{.15\textwidth}p{.15\textwidth}p{.15\textwidth}}\hline
\textbf{Variables} & \multicolumn{4}{c}{\textbf{Principle Components and Factor Loadings}}\\
\hline
 &$\mathbf{Wikipedia_{1}}$ &$\mathbf{Wikipedia_{2}}$ & & \\
\hspace*{.5em}$W_{purpose}$ &0.71 &-0.71 & & \\
\hspace*{.5em}$W_{frequency}$ &0.71 &0.71 & & \\[.5cm]
 &$\mathbf{Computer_{1}}$ &$\mathbf{Computer_{2}}$ & & \\
\hspace*{.5em}$C_{frequency}$ &0.58 &-0.53 & & \\
\hspace*{.5em}$C_{good}$ &0.54 &0.82 & & \\
\hspace*{.5em}$C_{like}$ &0.61 &-0.21 & & \\[.5cm]
 &$\mathbf{Spatial_{1}}$ &$\mathbf{Spatial_{2}}$ &$\mathbf{Spatial_{3}}$ &$\mathbf{Spatial_{4}}$ \\
\hspace*{.5em}$S_{good}$ &0.58 &-0.13 &-0.22 &-0.44 \\
\hspace*{.5em}$S_{learn}$ &0.44 &-0.48 &-0.30 &0.69 \\
\hspace*{.5em}$S_{unknown}$ &0.59 &0.19 &0.10 &-0.33 \\
\hspace*{.5em}$S_{known}$ &0.33 &0.61 &0.46 &0.47 \\
\hspace*{.5em}$S_{left}$ &0.06 &-0.59 &0.80 &-0.08 \\[.5cm]
 &$\mathbf{Education_{1}}$ &$\mathbf{Education_{2}}$ & & \\
\hspace*{.5em}$ED_{years}$ &0.71 &-0.71 & & \\
\hspace*{.5em}$ED_{highest}$ &0.71 &0.71 & & \\[.5cm]
 &$\mathbf{Employment_{1}}$ &$\mathbf{Employment_{2}}$ &$\mathbf{Employment_{3}}$ & \\
\hspace*{.5em}$EM_{status}$ &0.30 &-0.55 &0.77 & \\
\hspace*{.5em}$EM_{mental}$ &0.60 &-0.06 &-0.18 & \\
\hspace*{.5em}$EM_{physical}$ &0.06 &0.74 &0.56 & \\
\hspace*{.5em}$EM_{intensive}$ &0.54 &0.37 &-0.08 & \\
\hspace*{.5em}$EM_{creative}$ &0.51 &-0.09 &-0.22 & \\[.5cm]
\hline
\end{tabular}
\end{table}

\clearpage

\section*{Acknowledgements}

We are grateful to Csaba Pleh, Peter Kardos, and Markus Strohmaier for their valuable advice. This project was supported by the Humboldt Foundation within the Research Group Linkage Program. JK and MZ were partially supported through ERC grant No. 810115-DYNASET. MZ acknowledges further support from 101086712-LearnData-HORIZON-WIDERA-2022-TALENTS-01 financed by EUROPEAN RESEARCH EXECUTIVE AGENCY (REA), CORDIS. JK acknowledges further support from EU H2020 ICT48 project "Humane AI Net", grant No. 952026, and Horizon 2020 "INFRAIA-01-2018-2019" project "SoBigData++", grant No. 871042.

\section*{Author contributions statement}

All authors contributed to the conception and design of the research. MZ led the experiment and collected the data. All authors analyzed the data and wrote the paper.

\section*{Data availability}

Raw data for the online experiment has restricted access and can be provided upon consultation. Request for data should be directed to the corresponding authors.

\section*{Competing interests}

The authors declare no competing interests.

\section*{Ethics declarations}

All subjects gave their informed consent for inclusion before they participated in the study. The protocol of the study was approved by the Ethics Committee of Central European University (reference number: 2022-2023/1/EX). All methods of the study were carried out following the principles of the Belmont Report. 

\section*{Supplementary information}
The supplementary materials are included.

\end{document}